\documentclass[aps, prr, twocolumn,showkeys,showpacs,amsmath,amssymb,superscriptaddress,notitlepage,floatfix]{revtex4-1}

\usepackage{graphicx,graphics}
\usepackage{dcolumn}
\usepackage{float}
\usepackage{amsmath,amssymb,amsfonts}
\usepackage{latexsym,verbatim}
\usepackage[normalem]{ulem}
\usepackage{bm,bbold}
\usepackage{hhline}
\usepackage{enumerate}
\usepackage{color}
\usepackage{ulem}
\usepackage[percent]{overpic}
\usepackage[breaklinks=true,colorlinks,citecolor=blue,linkcolor=blue,urlcolor=blue]{hyperref}
\usepackage[utf8]{inputenc}
\usepackage{pgfplots}
\usepackage{subcaption}

\DeclareUnicodeCharacter{2212}{−}
\usepgfplotslibrary{groupplots,dateplot}
\usetikzlibrary{patterns,shapes.arrows}
\pgfplotsset{compat=newest}

\definecolor{ForestGreen}{RGB}{34,139,34}

\begin{document}

\title{Kinetic data-driven approach to turbulence subgrid modeling}

\author{G. Ortali}
\affiliation{Department of Applied Physics and Science Education, Eindhoven University of Technology, 5600 MB Eindhoven, The Netherlands}
\affiliation{SISSA (International School for Advanced Studies), Trieste, Italy}
\author{A. Gabbana}
\affiliation{Department of Applied Physics and Science Education, Eindhoven University of Technology, 5600 MB Eindhoven, The Netherlands}
\affiliation{Eindhoven Artificial Intelligence Systems Institute, Eindhoven University of Technology, 5600 MB Eindhoven, The Netherlands}
\author{N. Demo}
\affiliation{SISSA (International School for Advanced Studies), Trieste, Italy}
\author{G. Rozza}
\affiliation{SISSA (International School for Advanced Studies), Trieste, Italy}

\author{F. Toschi}
\affiliation{Department of Applied Physics and Science Education, Eindhoven University of Technology, 5600 MB Eindhoven, The Netherlands}
\affiliation{Eindhoven Artificial Intelligence Systems Institute, Eindhoven University of Technology, 5600 MB Eindhoven, The Netherlands}
\affiliation{CNR-IAC, I-00185 Rome, Italy}

\begin{abstract}
  Numerical simulations of turbulent flows are well known to pose extreme computational challenges due to the huge number of dynamical degrees of freedom required to correctly describe the complex multi-scale statistical correlations of the velocity. On the other hand, kinetic mesoscale approaches based on the Boltzmann equation, have the potential to describe a broad range of flows, stretching well beyond the special case of gases close to equilibrium, which results in the ordinary Navier-Stokes dynamics. 
  Here we demonstrate that, by properly tuning, a kinetic approach can statistically reproduce the quantitative dynamics of the larger scales in turbulence, thereby providing an alternative, computationally efficient and physically rooted approach towards subgrid scale (SGS) modeling in turbulence.
  More specifically we show that by leveraging on data from fully resolved Direct Numerical Simulation (DNS) we can learn a collision operator for the discretized Boltzmann equation solver (the lattice Boltzmann method), which effectively implies a turbulence subgrid closure model.
  The mesoscopic nature of our formulation makes the learning problem fully local in both space and time, leading to reduced computational costs and enhanced generalization capabilities. 
  We show that the model offers superior performance compared to traditional methods, such as the Smagorinsky model, being less dissipative and, therefore, being able to more closely capture the intermittency of higher-order velocity correlations.
  This foundational work lays the basis for extending the proposed framework to different turbulent flow settings and -most importantly- to develop new classes of hybrid data-driven kinetic-based models capable of faithfully capturing the complex macroscopic dynamics of diverse physical systems such as emulsions, non-Newtonian fluid and multiphase systems.
\end{abstract}

\maketitle

{\it Introduction.---}
Studying turbulent flows by means of fully resolved numerical simulations is known to pose outstanding computational challenges. The number of dynamical degrees of freedom, whose dynamics needs to be accurately resolved, is typically huge as it rapidly grows as the $9/4$ power of the Reynolds number ($\rm{Re}$)~\cite{frisch-book-1995,pope-book-2001,benzi-pr-2023}.
Despite the ever increasing availability of computing resources, Direct Numerical Simulations (DNS)
of turbulent flows are still far out of reach for most real-world applications, strongly motivating a continuous effort towards the development of accurate and computationally cheaper reduced order models.
Alongside Reynolds-Averaged Navier-Stokes (RANS) models~\cite{alfonsi-amr-2009}, where the Navier-Stokes equations
are averaged over time (separating the flow into mean and fluctuating components) 
large-eddy simulations (LES)~\cite{zhiyin-cja-2015} are among the most popular choices.
In LES only a portion of the dynamical degrees of freedom, associated to the larger scales, are directly resolved,  whereas the effect of the smaller scales on the large scales is parametrized by a subgrid-scale (SGS) model. 
In general, considering the filtered version of the macroscopic fields of interest, $\overline{\bm{u}}$ and $\overline{p}$ (respectively, velocity and pressure), 
one can write down the Navier-Stokes equations filtered to describe only scales larger than $\ell$:
\begin{equation}\label{eq:filtered-nse}
    \frac{\partial \overline{\bm{u}}}{\partial t} 
  + \nabla \cdot (\overline{\bm{u}}~\overline{\bm{u}} + \overline{\bm{T}}) 
  = 
  - \nabla \overline{p}  
  + \nu \nabla^2 \overline{\bm{u}} .
\end{equation}
In this filtered equation, we have assumed the density $\rho=1$, $\nu$ is the kinematic viscosity of the fluid and
$\overline{\bm{T}} = \overline{\bm{u} \bm{u}} - \overline{\bm{u}}~\overline{\bm{u}}$ 
is the Reynolds stress tensor, accounting for the SGS fluctuations, from which one can define a local subgrid energy flux \cite{borue-jfm-1998}.
The problem of defining an SGS closure consists in modelling the unknown term $\overline{\bm{T}}$ 
as a function of only the resolved velocity field $\overline{\bm{u}}$. 
Over the years, several approaches to SGS turbulence modeling have been developed, relying on specific
assumptions to approximate the effects of the unresolved scales. 
\begin{figure*}[htb]
  \centering
  \begin{overpic}[width=.99\textwidth]{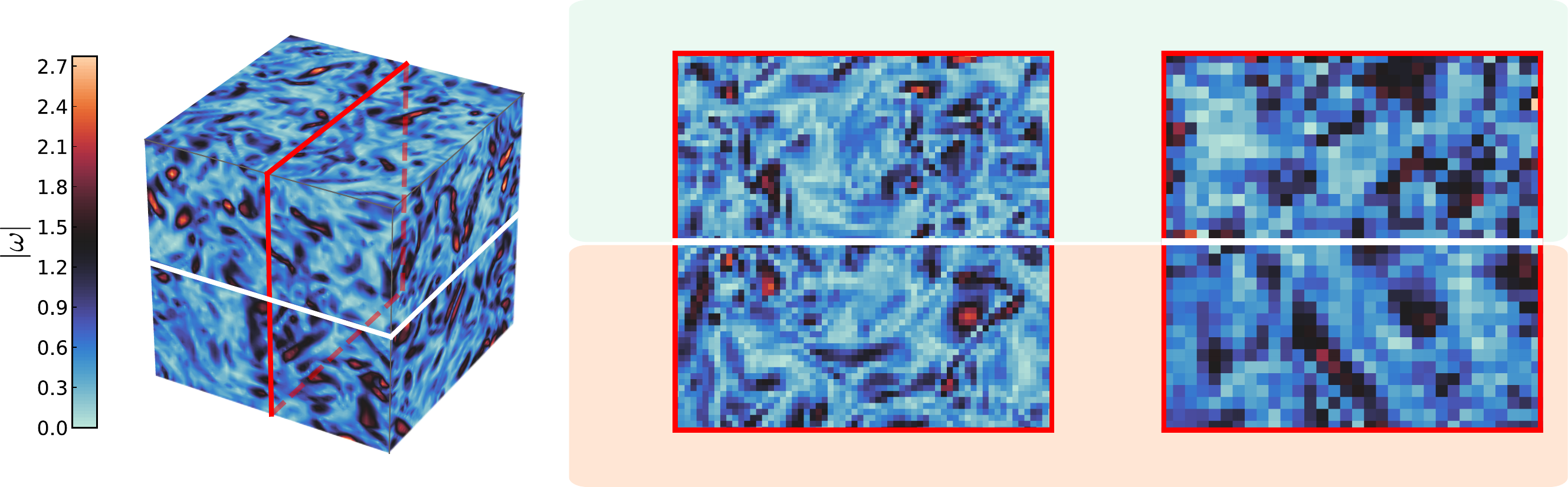}
    \put(0 ,30){(a)}
    \put(32,30){(b)}
    \put(70,30){(c)}
    \put(37,29.5){\color{black} \textbf{FILTERED DNS }}
    \put(37, 0.5){\color{black} \textbf{NEURAL LBM}}    
    \put(61,29){\normalsize  $\rm{cg} = 2$}
    \put(92,29){\normalsize  $\rm{cg} = 4$}
  \end{overpic}
  \caption{Snapshots of the vorticity magnitude ($|\omega|$) from 3D simulations of HIT at $\rm{Re} \approx 6000$.
           In panel a) the upper half of the domain is taken from filtered DNS data with $\rm{cg} = 2$, while 
           the lower half is obtained from a simulation using NBLM, trained and validated on the same flow conditions, albeit with different initial configurations. 
           We show that at a qualitative level the structures generated 
           by NLBM closely resemble those from filtered DNS. This is further highlighted in panel b) and c)
           where we show 2D slices of $|\omega|$ (cf. red box in panel a) ), respectively at coarse graining 
           factor $\rm{cg} = 2$ and $\rm{cg} = 4$.
           }\label{fig:slices}
\end{figure*}
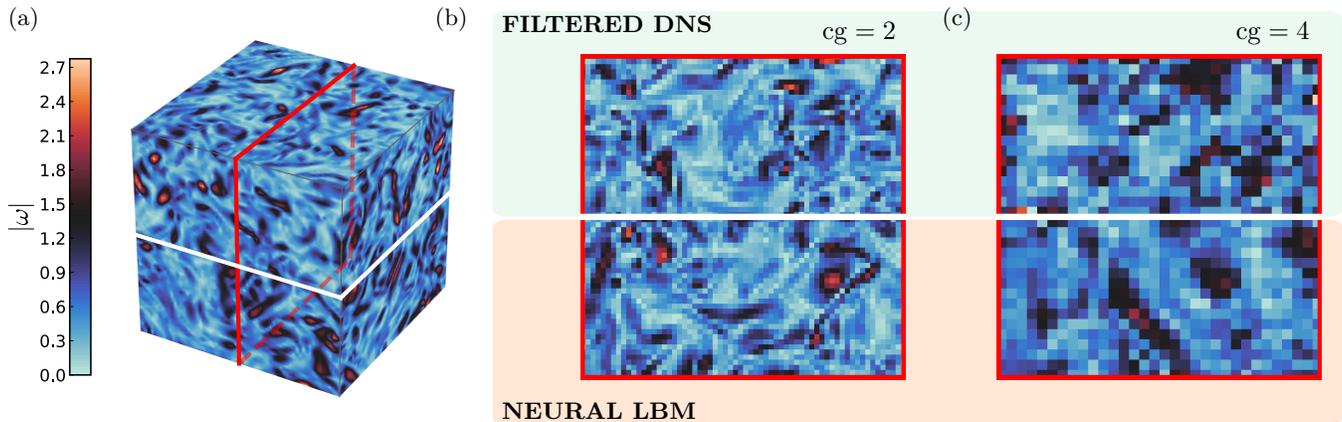
The first SGS model was proposed by Smagorinsky~\cite{smagorinsky-mwr-1963}
and, in view of its simplicity and stability, it is one of the most widespread
adopted approaches to estimate the action of smaller scales on the larger scales by means of an effective 
eddy viscosity, defined on the basis of the local derivatives of the resolved velocity field:
    \begin{equation}\label{eq:smago-sgs}
    \overline{\bm{T}}^{\rm smag} = 2 \nu_{\rm t} \overline{\bm{S}} - \overline{p} \bm{I} , 
    \quad 
    \nu_{\rm t} = (C_s \Delta)^2 |\overline{\bm{S}}| ,
    \end{equation}

where $|\overline{\bm{S}}| = \sqrt{2 \overline{S}_{ij} \overline{S}_{ij}}, \overline{S}_{ij} = 1/2 (\partial_i \overline{u}_{j} + \partial_j \overline{u}_{i} )$ 
represents the magnitude of the filtered strain-rate, $\Delta$ is the filter width, and $C_s$ is a 
non-dimensional coefficient called the Smagorinsky constant.
The Smagorinsky model can be shown to be a direct consequence of the Refined Kolmogorov Similarity Hypothesis~\cite{toschi-prl-2000} 
and has been generalized to non homogeneous flows~\cite{leveque-jfm-2007}.

Other approaches have been introduced, including improvements on the Smagorinsky model that dynamically adjust the Smagorinsky constant $C_s$ based on the flow field~\cite{germano-pof-1991,ghosal-jfm-1995}, and scale similarity models~\cite{meneveau-arfm-2000}. 
However, despite their widespread use, these models often face limitations in accurately representing the (statistical) 
properties of turbulent flows, particularly in scenarios with strong inhomogeneity and anisotropy.

Recent advances in Machine Learning have opened up new perspectives for employing Artificial Neural Networks (ANN)~\cite{goodfellow-book-2016} to enhance computational fluid dynamic solvers~\cite{kochkov-pnas-2021,vinuesa-ncs-2022,lino-prsa-2023,brunton-arfd-2020}, and for the development of 
data-driven turbulence modelling~\cite{novati2021automating,ling-jfm-2016, duraisamy-arfm-2019, wang-sigkdd-2020}.
Specifically to the LES context, several attempts have been reported in the literature~\cite{wang-pof-2018,maulik-jfm-2019, xie-prf-2020,
park-jfm-2021, frezat-prf-2021, tian-pnas-2023} to establish SGS closure models from extensive datasets of fully resolved turbulent flows, 
leveraging on the capability of ANN to handle high-dimensional and statistically complex data.
In general, two main approaches have been adopted; the first, which can be regarded as the "black-box" approach, 
consists in making no assumption on the form of the SGS terms~\cite{beck-jcp-2019,zhou-cf-2019,xie-pre-2019,kurz-ijhff-2023}, 
while in the second one the task of the ANN consists in tuning the free-parameters of an already established 
SGS model~\cite{sarghini-cf-2003,xie-pof-2019,wang-aip-2021}.

In this Letter we take a fresh look at the problem of establishing a SGS closure model with ANN, and 
introduce for the first time, to the best of our knowledge, a data-driven kinetic-based approach to turbulence modeling.
We employ Physics-Informed Machine Learning (PIML) to enhance the capabilities of the
Lattice Boltzmann Method (LBM), and exploit the 
extra degrees of freedom provided by  
the mesoscopic description to learn a new collision operator which effectively acts as SGS model. 
Our framework relies on the pathway %
connecting kinetic theory to hydrodynamics, offering the possibility 
to learn from data macroscopic equations that extend beyond the Navier-Stokes level.
Specifically, we consider the problem of learning an SGS closure in the context of homogeneous
isotropic turbulence (HIT) that reproduces the complex statistics of filtered DNS, including
features such as intermittency and backscatter. This represents an open problem in turbulence 
research and is considered here as a foundational step for the introduced framework, which could be 
extended to learn different physical problems.
The proposed framework is inherently local due to the locality of the collision operator~\cite{chen-sc-2003}, which contrasts with other ML approaches~\cite{ling-jfm-2016}. 
This drastically simplifies the training process and enhances the interpretability of the model.
Remarkably, and at variance with respect to other fully local SGS models, our formalism shows the 
potential for capturing the inverse transfer of energy from small to large scales,  without causing numerical instabilities.

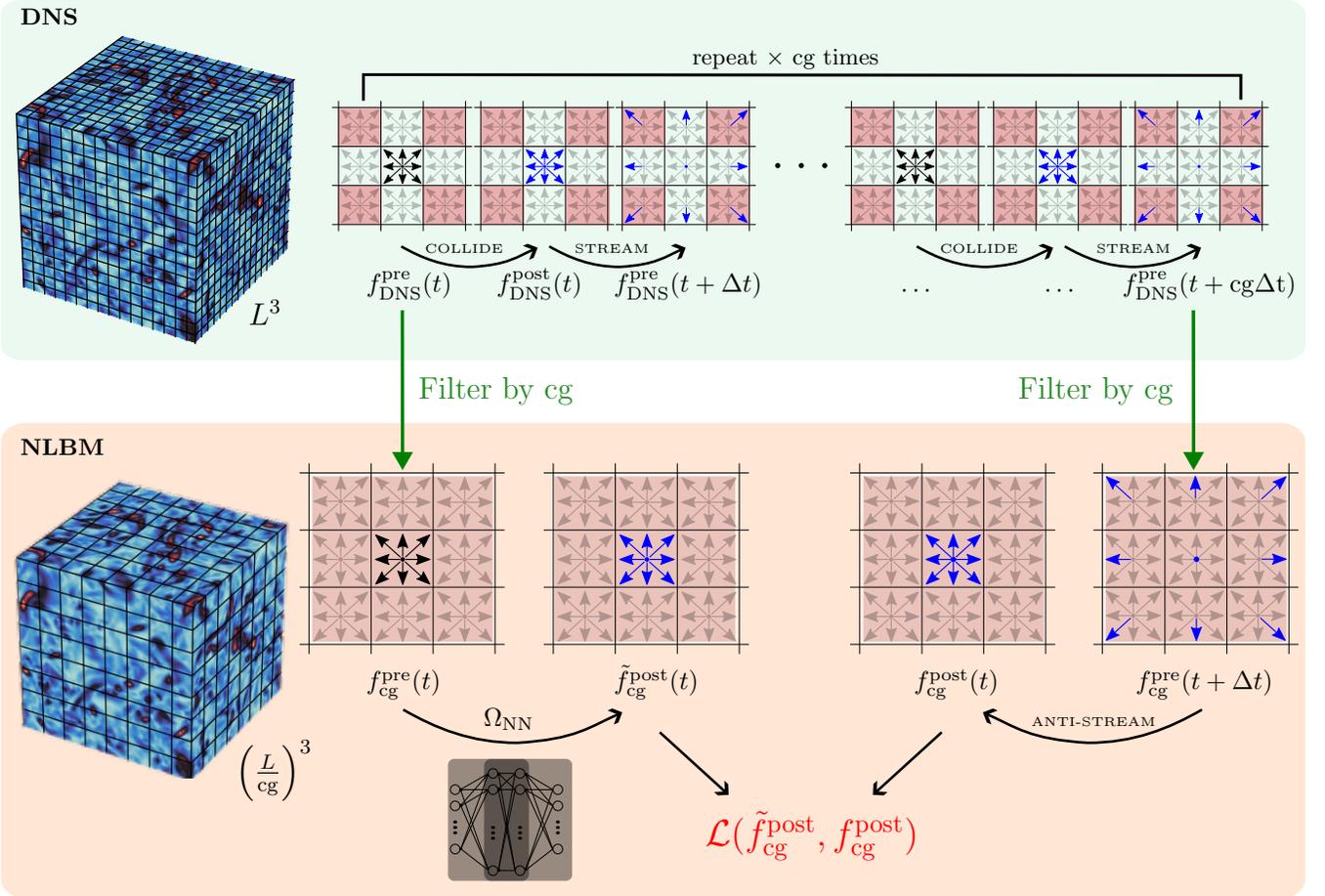
\begin{figure*}[htb]
  \centering
  \begin{overpic}[width=.99\textwidth]{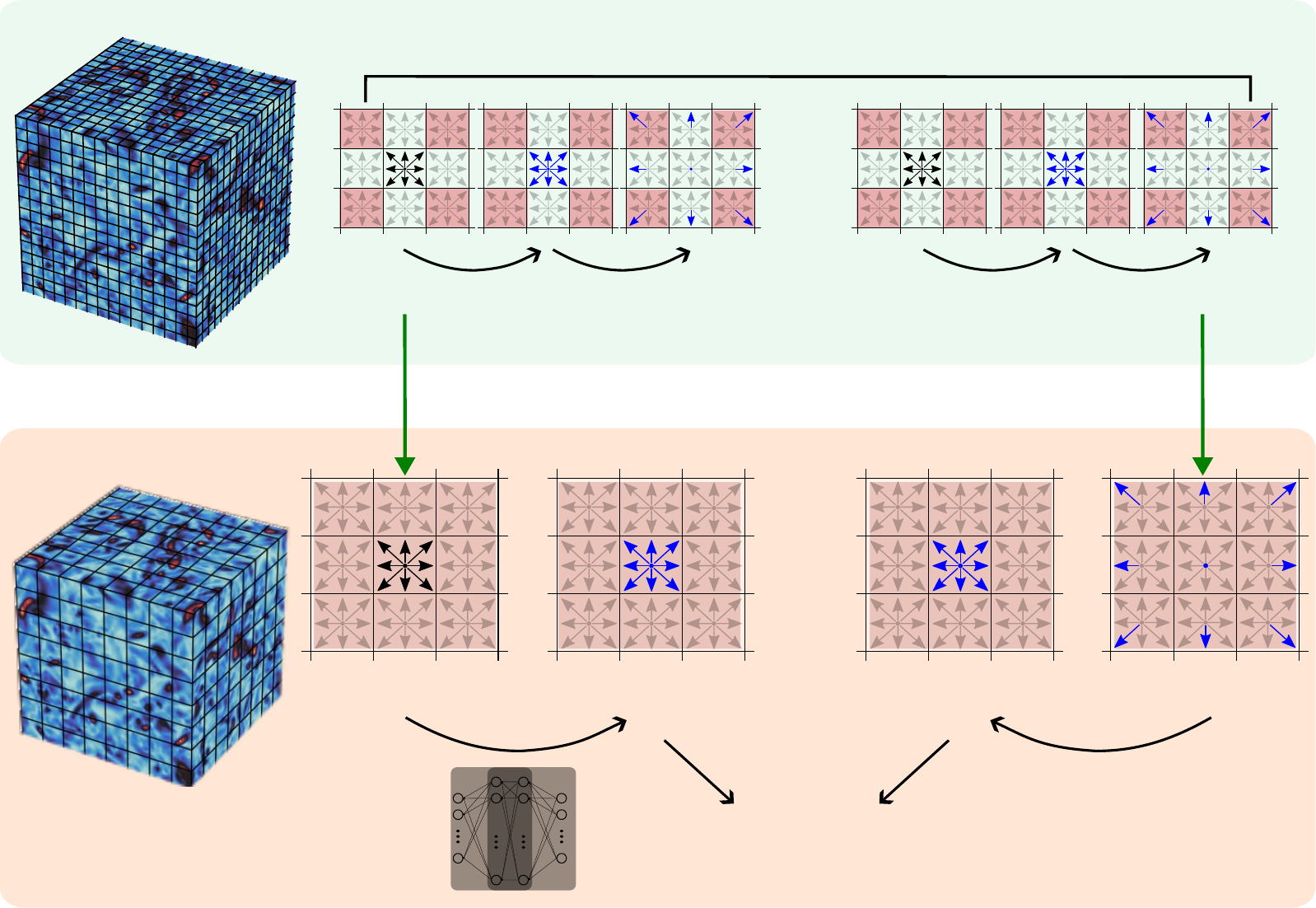}
    \put(19,44){\large $L^3$}
    \put(18, 9){\large $\left( \frac{L}{\rm{cg}} \right)^3$}
    \put(1.5, 67){\color{black} \textbf{DNS  }}
    \put(1.5, 34){\color{black} \textbf{NLBM }}
    \put(53, 64){repeat $\times$ cg times}
    \put(32, 38.3){\large \color{ForestGreen} Filter by $\rm cg$}
    \put(78, 38.3){\large \color{ForestGreen} Filter by $\rm cg$}
    \put(59, 56){\huge $\dots$}
    \put(28, 46.5){\normalsize $      {f}^{\rm pre  }_{\rm DNS }(t           )$}
    \put(38, 46.5){\normalsize $      {f}^{\rm post }_{\rm DNS }(t           )$}
    \put(47, 46.5){\normalsize $      {f}^{\rm pre  }_{\rm DNS }(t + \Delta t)$}
    \put(69, 46.5){\normalsize $\dots$}    
    \put(80, 46.5){\normalsize $\dots$}    
    \put(86, 46.5){\normalsize $      {f}^{\rm pre  }_{\rm DNS }(t + \rm{cg} \Delta t)$}    
    \put(32.5, 49.5){\tiny COLLIDE}
    \put(44.0, 49.5){\tiny STREAM }
    \put(72.0, 49.5){\tiny COLLIDE}
    \put(84.0, 49.5){\tiny STREAM }    
    \put(28, 16.0){\normalsize $      {f}^{\rm pre }_{\rm cg }(t           )$}
    \put(47, 16.0){\normalsize $\tilde{f}^{\rm post}_{\rm cg }(t           )$}
    \put(70, 16.0){\normalsize $      {f}^{\rm post}_{\rm cg }(t           )$}
    \put(87, 16.0){\normalsize $      {f}^{\rm pre }_{\rm cg }(t + \Delta t)$}
    \put(37, 13.2){\normalsize $ \Omega_{\rm NN }$}
    \put(79, 13.2){\tiny ANTI-STREAM}
    \put(54, 4.0){\Large \color{red} $ \mathcal{L}(\tilde{f}^{\rm post}_{\rm cg },  f^{\rm post}_{\rm cg })  $}
  \end{overpic}  
  \caption{Schematic representation of the training process for the turbulence SGS model. 
           The upper panel corresponds to the DNS simulation on a $L^3$ grid, while in lower panel the Neural LBM (NLBM)
           operates on a $(L/\rm{cg})^3$ grid, with coarse-graining factor $\rm{cg}$ via subsampling.
           The mapping between DNS into coarse-grained data is given by the application of a filter:
           the pre-collision state at a generic time step $t$ at the coarse grained level ( $f_{\rm cg}^{\rm pre}(t)$ ) 
           is obtained by filtering the DNS pre-collision state ( $f_{\rm DNS}^{pre}(t)$ ) (dependence on space
has been omitted for conciseness).
           Similarly, the post-collision data at the coarse-grained level ( $f_{\rm cg}^{\rm post}(t)$ ) is obtained 
           by first filtering the post-collision DNS state at time step $t + \rm{cg} \Delta t$, and then by 
           applying the inverse of the streaming operator.
           Following this procedure it is possible to create a dataset of arbitrary size for training an ANN to which
           we assign the task of minimizing the mismatch between 
           $\tilde{f}_{\rm cg}^{\rm post}(t) = \Omega_{\rm NN}(f_{\rm cg}^{\rm pre}(t))$
           and $f_{\rm cg}^{\rm post}(t)$ under a given error-metric $\mathcal{L}$.
           }\label{fig:training-diagram}
\end{figure*}

{\it Methods.---} 
The Lattice Boltzmann Method (LBM)~\cite{benzi-pr-1992,succi-book-2018,kruger-book-2017} has emerged in the past 
decades as a popular solver for computational fluid dynamics.
At variance with methods that explicitly discretize the Navies-Stokes equations, LBM operates at the mesoscopic level, 
providing the description of a fluid system in terms of a (small) set of particle distribution functions (populations)
whose dynamic is governed by the discrete Boltzmann equation:
\begin{equation}\label{eq:lbe}
  f_i( \bm{x} + \bm{c}_i \Delta t, t + \Delta t) = f_i(\bm{x}, t) + \Omega_i(\bm{x}, t) ,
\end{equation}
where at each grid node $\bm{x}$, lattice populations $f_i (\bm{x},t)$ represent the probability distribution function of particles at position $\bm{x}$ and time $t$ moving with discrete velocity $\bm{c}_i, i = 1, \dots , Q$. The uninitiated reader can find a more detailed introduction to the Lattice Boltzmann Method in the Supplementary Information.
A popular choice for the collision operator $\Omega$ is the single-time relaxation BGK model~\cite{bhatnagar-pr-1954}:
\begin{equation}\label{eq:lbgk}
  \Omega_i(\bm{x},t)
  = 
  -\frac{\Delta t}{\tau} \left( f_i(\bm{x},t) - f_{i}^{\rm{eq}} (\bm{x},t) \right),
\end{equation}
which models collisions as a relaxation towards an equilibrium distribution,
with $\tau$ the relaxation rate.
The macroscopic variables of interest can be obtained as the lower-order moments
of the lattice populations. Moreover, it can be shown via an asymptotic analysis that Eq.~\ref{eq:lbe} yields 
a second order approximation of the Navier-Stokes equations~\cite{kruger-book-2017}.

It is an expedient for the description of our method to split 
the time evolution of Eq.~\ref{eq:lbe} into two steps, 
following the \textit{stream} and \textit{collide} paradigm:
\begin{align}
  f_i^{\rm post}(\bm{x}, t          ) &= f_i^{\rm pre}(\bm{x},t) + \Omega_i(\bm{x}, t), \\
  f_i^{\rm pre }(\bm{x}, t+ \Delta t) &= f^{\rm post}_i(\bm{x} - \bm{c}_{i} \Delta t , t),  
\end{align}
where here and in what follows we will denote with $f_i^{\rm pre}$ ($f_i^{\rm post}$) the pre(post)-collision populations.

Following the framework introduced in Ref.~\cite{corbetta-epje-2023}, we replace the collision operator with a ANN, effectively defining a generalized collision operator:
\begin{equation}\label{nn-model}
  \tilde{f}_i^{\rm post}(\bm{x}, t) = f_i^{\rm pre}(\bm{x},t) + \Omega^{\rm NN}\left( f_i^{\rm pre}(\bm{x},t) \right) .
\end{equation}
The resulting algorithm, to which we refer as Neural Lattice Boltzmann Method (NLBM), employs a physics-constrained 
ANN to establish a data-driven SGS model. Our approach leverages on two key ingredients of the LBM algorithm:
\begin{enumerate}
  \item The LBM mesoscopic representation makes use of a larger number of degrees of freedom 
        (i.e. the number of discrete lattice populations) than the macroscopic observable of interest.
        This observation opens up the possibility of using ANN to encode extra information in the model.
  \item The non-linear terms encountered at the NS level  of description (cf. Eq.~\ref{eq:filtered-nse}) are 
        fully embedded in the LBM via the collision operator and thus purely local 
        (in LBM ``non-linearity is local, non-locality is linear''~\cite{succi-book-2018}).
        This observation drastically reduces the cost of training ANN, since it offers the possibility to restrict
        the input and output of the network to local quantities without explicit evaluation of gradients of 
        macroscopic fields.
\end{enumerate}

We summarize here the main steps required to train the ANN, as sketched in Fig.~\ref{fig:training-diagram}, 
leaving full technical details to the Supplementary Material.
We consider simulations of Homogeneous Isotropic Turbulence (HIT), fully resolved on a $L^3$ domain using the 
standard LBM formulation (Eq.~\ref{eq:lbe}) with BGK collision operator (Eq.~\ref{eq:lbgk}).
We regard this as the target ground truth DNS data (upper panel in Fig.~\ref{fig:training-diagram}).
Next, we define a coarse-graining factor $\rm{cg}$, and define a filter which projects, at a given time step $t$,
data from the DNS grid to a coarse-grained grid of size $(L / \rm{cg})^3$, where the SGS model will operate 
(lower panel in Fig.~\ref{fig:training-diagram}).
In order to define an arbitrarily large training dataset consisting of pre and post collision data at the coarse-grained level,
respectively $f_{\rm cg}^{\rm pre}(t)$ and $f_{\rm cg}^{\rm post}(t)$, we take the following steps:
We start from the pre-collision populations at the DNS level, $f_{\rm DNS}^{\rm pre}(t)$, and immediately apply the filter
in order to obtain $f_{\rm cg}^{\rm pre}(t)$. Next, we advance the DNS simulation for $\rm{cg} \Delta t$ time-steps,
yielding the pre-collision (and post-streaming) populations $f_{\rm DNS}^{\rm pre}(t + \rm{cg} \Delta t)$.
Since $\Delta t$ steps at the coarse-grained level correspond to $\rm{cg} \Delta t$ steps at the DNS level,
by filtering $f_{\rm DNS}^{\rm pre}(t + \rm{cg} \Delta t)$ we obtain $f_{\rm cg}^{\rm pre}(t + \Delta t)$. 
Finally, we obtain $f_{\rm cg}^{\rm post}(t)$ by reversing the streaming operation, 
i.e. we anti-stream populations on the coarse grid with respect to the corresponding velocity component $-\bm{c}_i$ taken from the coarse grid velocity stencil.

At this stage, we can train a ANN which, under a given error-metric $\mathcal{L}$, minimizes the mismatch 
between $f_{\rm cg}^{\rm post}(t)$ and the prediction of the network taking $f_{\rm cg}^{\rm pre}(t)$ as input:
\begin{equation}
  \mathcal{L} \left( \Omega^{\rm NN}(f_{\rm cg}^{\rm pre}(t)), f_{\rm cg}^{\rm post}(t) \right).
\end{equation}

We have observed that the two main ingredients allowing for training models which deliver accurate and stable 
results throughout an entire simulation are i) imposing hard-constraints on the preservation of mass and momentum~\cite{corbetta-epje-2023}
and ii) using \textit{unrolled training}~\cite{brandstetter-arxiv-2022} to compute the loss over consecutive timesteps.
Full details on the ANN architecture and the training process and the complete form of
the loss function are provided as Supplementary Information.

\begin{figure}[htb]
  \centering
  \includegraphics[width=0.99\columnwidth]{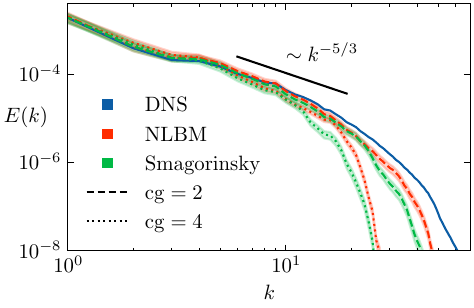}
  \caption{ Energy spectrum for simulations of HIT at $\rm{Re} \approx 6000$.
            The results from DNS (blue curve) are compared with NLBM (red)
            and Smagorinsky (green). For the two SGS models we report the average
            spectrum from $80$ simulations starting from different initial conditions.
            The shaded curves corresponds to one standard deviation from the average value.
          }\label{fig:spectra}
\end{figure}

{\it Numerical results.---} 
We consider numerical simulations of HIT, with for reference a DNS on a $L = 128^3$ grid at Reynolds number 
$\rm{Re} \approx 6000$ ( corresponding to a Taylor microscale Reynolds Number $\rm{Re}_{\lambda} \approx 77$ ). 
The parameters are selected in such a way that the plain LBM algorithm would encounter 
numerical instabilities when working on coarse-grained grids with $\rm{cg} = \{ 2, 4 \}$.
Fig.~\ref{fig:slices} qualitatively summarizes our findings. Starting from the same initial configuration, we 
present 3D and 2D representation of the absolute value of the vorticity $| \bm{\omega} | = |\nabla \times \bm{u}|$
at a late stage of the simulation, comparing results from the filtered DNS with NLBM. Our model provides stable 
simulations, with flow patterns virtually indistinguishable from those of the filtered DNS.

\begin{figure}[htb]
  \centering
  \includegraphics[width=0.99\columnwidth]{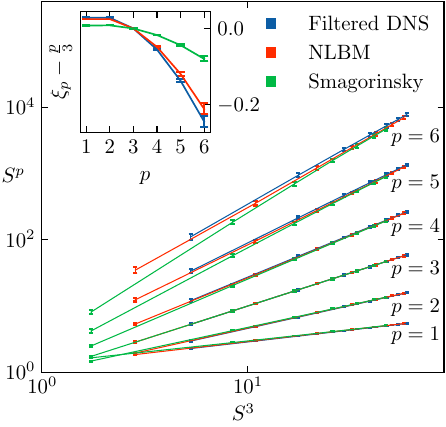}
  \caption{Structure functions (cf. Eq.~\ref{eq:structure-fun}) of order $p$, ranging between $p = 1$ to $p = 6$, 
           versus $S^3$, with in blue data from filtered DNS with $\rm{cg} = 4$, in red data from simulations using NLBM, 
           and in green data from simulations using the Smagorinsky model.
           The inset shows the deviation for the scaling exponents $\xi_p$ from the K41 scaling $\frac{p}{3}$.
         }\label{fig:ess}
\end{figure}

On a more quantitative ground, we now turn to the evaluation of the statistical properties of the turbulent flows
produced by NLBM simulations.
For comparison, we include LBM simulations equipped with the Smagorinsky SGS
model (Eq.~\ref{eq:smago-sgs}) as a standard reference point. The Smagorinsky model is chosen in this
context for its well-known and widely understood properties within the context of HIT.
In Fig.~\ref{fig:spectra} we present the kinetic energy spectrum $E(k)$. We observe that NLBM well compares with
the results produced by Smagorinsky, with slightly superior results at $\rm{cg} = 4$, where we can observe a 
spectrum less dissipative than the Smagorinsky one, and closer to the scaling of DNS data.
Next, we study the scaling behaviour of high-order Eulerian Structure function, which for a generic 
order $p$ can be computed from numerical data as
\begin{equation}\label{eq:structure-fun}
  S^p(l) = \left\langle \left[\left( \bm{u}(\bm{x} + \bm{l}) - \bm{u}(\bm{x}) \right)\cdot {\hat{\bm{l}}}\right]^p \right\rangle .
\end{equation}
We use the extended self similarity (ESS)~\cite{benzi-pre-1993} to determine the scaling exponents $\xi_p$.
In Fig.~\ref{fig:ess} we plot the structure functions of order $p$, ranging between $p = 1$ to $p = 6$, 
versus $S^3(l)$, comparing data from filtered DNS with $\rm{cg} = 4$, and data from simulations using NLBM and 
the Smagorinsky model.
In the inset of Fig.~\ref{fig:ess} we report the deviation of $\xi_p$ from the K41 scaling ${p}/{3}$.
The results highlight that, due to its lower dissipation at small scales, NLBM provides anomalous scaling 
within error bars from filtered DNS. On the other hand, the discrepancies observed in the Smagorinsky model 
are due to the fact that under these parameters the inertial range is shrinked (see the supplementary material
for a more detailed analysis).

\begin{figure}[htb]
  \centering
  \includegraphics[width=0.99\columnwidth]{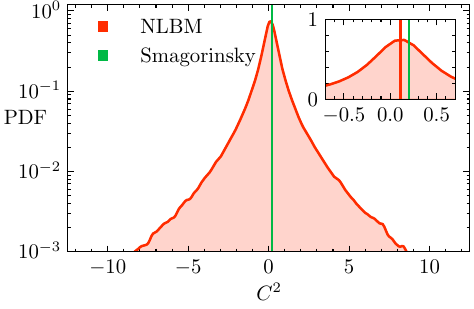}
  \caption{
            Probability distribution function (PDF) of the fitted
            value of the Smagorinsky constant $C^2$ from NLBM data.
            The inset highlights that the average value 
            $<C^2_{\rm NLBM}> \approx 0.11$ is about a factor two
            smaller than the one used in simulations with the 
            Smagorinsky SGS model ($C^2 = 0.2$).
            The presence of a tail with negative values highlights 
            the fact that in NLBM it is possible to capture 
            the inverse transfer of energy from small to large scales.
          }\label{fig:nu-eff}
\end{figure}

A major advantage of our kinetic-based approach lies in the possibility of having a physical interpretation of the action 
of the ANN, by projecting the lattice populations to the velocity moment space (see Supplementary Material).
We have observed that NLBM extends the single-relaxation BGK collision operator, from where DNS data was generated,
to a multi-relaxation collision operator~\cite{lallemand-pre-2000,dhumieres-ptsa-2002}. The model introduces 
a non-linear relation for pre and post collision values of the moments related to the bulk-viscosity,
something which is often used to increase the stability of numerical simulations. The model preserves a linear-dependency for 
the pre and post collision values of the moments related to the kinetic viscosity. 
This allows to fit the effective viscosity from the numerical data, which can be used to establish a direct comparison 
with the Smagorinsky model (Eq.~\ref{eq:smago-sgs}).
In Fig.~\ref{fig:nu-eff} we compare the probability distribution function (PDF) of the value of the Smagorinsky constant $C^2$ 
fitted from NLBM data (cg = 4). The average value of the Smagorinsky constant for NLBM ($ \langle C^2_{\rm NLBM} \rangle \approx 0.11$) 
is about a factor two smaller than the one used in simulations with the Smagorinsky SGS model ($C^2 = 0.2$).
Remarkably, in Fig.~\ref{fig:nu-eff} we observe a tail taking negative values, suggesting that our model occasionally 
displays an inverse transfer of energy from small to large scales, a feature completely lacking to the Smagorinski model 
which is, by its own nature, fully dissipative.

{\it Conclusion.---} 
Summarizing, we have introduced a novel kinetic-based approach to SGS modeling, combining LBM with physics informed ANN.
Our model allows for stable simulations on coarse domains, offering the possibility of reducing the computational costs
of DNS, in turn preserving the statistical properties of turbulent flows under HIT settings.
The model compares well with Smagorinsky, and our results highlight a better agreement with DNS in terms 
of energy spectra and estimation of the anomalous scaling exponents.
Moreover, we have shown that the model in principle supports the possibility to describe the inverse transfer of energy 
from small to large scales, which we regard as  promising particularly in vision of future application to more involved numerical setups.
To conclude, this work opens up the possibility of exploiting the extra degrees of freedom of the mesoscopic representation
in the LBM to develop novel SGS models. In future works, we plan to extend our analysis to flows subject to anisotropy as well as wall-bounded flows.
A further intriguing question concerns the application of this framework beyond SGS. Work to explore employing our kinetic data-driven
approach in other contexts is currently underway.

{\it Acknowledgments.---}
We wish to thank Roberto Benzi for useful discussions.
This work was partially funded by the Dutch Research Council (NWO) through 
the UNRAVEL project (with Project No. OCENW.GROOT.2019.044).

\pagebreak
\widetext
\begin{center}
\textbf{\large Supplementary Information for \\``Kinetic data-driven approach to turbulence subgrid modeling''}
\end{center}
\setcounter{equation}{0}
\setcounter{figure}{0}
\setcounter{table}{0}
\setcounter{page}{1}
\makeatletter
\renewcommand{\theequation}{S\arabic{equation}}
\renewcommand{\thefigure}{S\arabic{figure}}
\renewcommand{\thetable}{S\arabic{table}}
\renewcommand{\bibnumfmt}[1]{[S#1]}
\renewcommand{\citenumfont}[1]{S#1}

\section{Numerical Method}
In this section, we provide a basic introduction to the Lattice Boltzmann Method (LBM). 
The reader is referred to Ref.~\cite{sm-kruger-book-2017, sm-succi-book-2018} for 
a thorough introduction to the topic.
The LBM is a class of numerical fluid-dynamics solvers. 
At variance with conventional methods that explicitly discretize the Navies-Stokes equations, 
LBM takes root from the mesoscopic Boltzmann equation:
\begin{equation}\label{eq:be}
    \frac{\partial f}{\partial t} 
  + \xi_{\alpha} \frac{\partial f}{\partial x_{\alpha}} 
  + F_{\alpha} \frac{\partial f}{\partial \xi_{\alpha}} 
  = 
  \Omega(f)
\end{equation}
where $f(\bm{x}, \bm{\xi}, t)$ is the particle distribution function, representing the average number 
of particles in a small element of phase-space centered at position $\bm{x}$ with velocity $\bm{\xi}$ at time $t$, 
and $F_{\alpha}$ is the sum of external forces acting on the system. 
The right hand side of the equation, $\Omega(f)$, is the collision operator, describing the changes 
in $f$ due to particle collisions.

\begin{minipage}{\textwidth}
  \begin{minipage}[b]{0.5\textwidth}
    \centering
    \includegraphics[width=1\columnwidth]{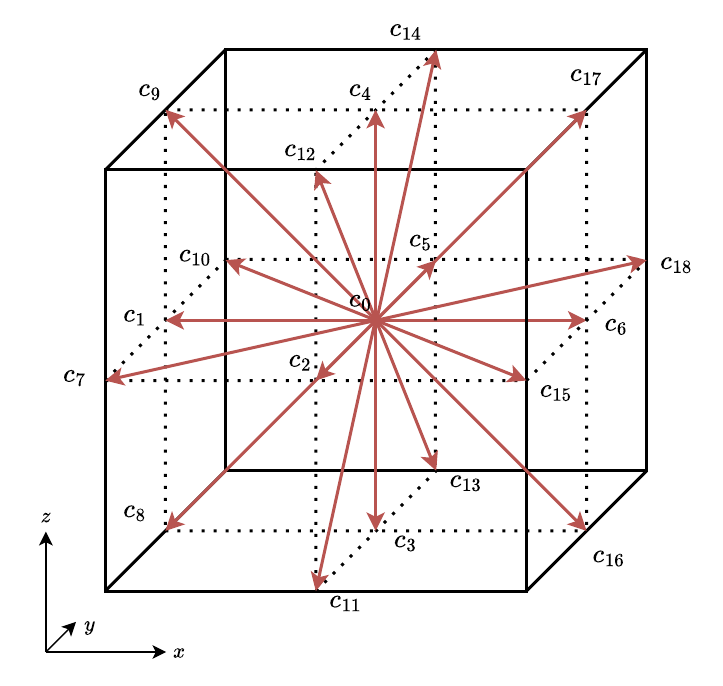}
    \captionof{figure}{D3Q19 Velocity stencil }
    \label{fig:d3q19}
  \end{minipage}
  \hfill
  \begin{minipage}[b]{0.49\textwidth}
    \centering
    \begin{tabular}{ccc}\hline
      Index & $\bm{c}_i$ & $w_i$ \\ \hline
                $0 $ &  $(0, 0, 0)$   & $1/3$ \\  
                $1 $ &  $(-1, 0, 0)$  & $1/18$ \\
                $2 $ &  $(0, -1, 0)$  & $1/18$ \\
                $3 $ &  $(0, 0, -1)$  & $1/18$ \\
                $4 $ &  $(0, 0, 1)$   & $1/18$ \\
                $5 $ &  $(0, 1, 0)$   & $1/18$ \\
                $6 $ &  $(1, 0, 0)$   & $1/18$ \\
                $7 $ &  $(-1, -1, 0)$ & $1/36$ \\
                $8 $ &  $(-1, 0, -1)$ & $1/36$ \\
                $9 $ &  $(-1, 0, 1)$  & $1/36$ \\
                $10$ &  $(-1, 1, 0)$  & $1/36$ \\
                $11$ &  $(0, -1, -1)$ & $1/36$ \\
                $12$ &  $(0, -1, 1)$  & $1/36$ \\
                $13$ &  $(0, 1, -1)$  & $1/36$ \\
                $14$ &  $(0, 1, 1)$   & $1/36$ \\
                $15$ &  $(1, -1, 0)$  & $1/36$ \\
                $16$ &  $(1, 0, -1)$  & $1/36$ \\
                $17$ &  $(1, 0, 1)$   & $1/36$ \\
                $18$ &  $(1, 1, 0)$   & $1/36$ \\
    \end{tabular}
    \captionof{table}{D3Q19 Velocity stencil - discrete velocity vectors $\bm{c}_i$ and corresponding weights $w_i$.}
    \label{tab:stencil-weights}
  \end{minipage}
  \vspace{0.5cm}
\end{minipage}

The lattice time-discrete counterpart of Eq.~\ref{eq:be} reads as 
\begin{equation}\label{eq:lbe-sm}
  f_i( \bm{x} + \bm{c}_i \Delta t, t + \Delta t) = f_i(\bm{x}, t) + \Omega_i(\bm{x}, t) + f_i^{\rm ext} ,
\end{equation}
In the above equation, known as lattice Boltzmann equation, the discrete particle distribution functions
(populations) $f_i(\bm{x}, t)$ move at the next time step $t + \Delta t$ with velocity $\bm{c}_i$ 
to a neighboring grid cell $\bm{x} + \bm{c}_i \Delta t$. Moreover, $f_i^{\rm ext}$ is the discrete version 
of the external force term.
The discretization of the velocity space is a key element in the definition of LBM; it consists in replacing the 
continuum velocity space with a small set of discrete velocities $ \mathcal{V} = \{ \bm{c}_i \in \mathbb R^3 \}$
corresponding to the abscissa of a Gauss-Hermite quadrature (with corresponding weights $w_i$ ) that ensure
that all moments of the distribution function up to a prescribed order are correctly recovered~\cite{sm-philippi-pre-2006, sm-shan-jcs-2016}.
In this work, we use the D3Q19 velocity stencil (Fig.~\ref{fig:d3q19}), defined in Tab.~\ref{tab:stencil-weights},
which allows to express mass and momentum as weighted sums of $f_i(\bm{x}, t)$:
\begin{equation}
  \rho = \sum_i f_i, \quad \rho \bm{u} = \sum_i f_i \bm{c}_i .
\end{equation}

There are several possible choices to approximate the collision operator $\Omega_i$. Arguably the simplest
one that can be used for Navier-Stokes simulations is the Bhatnagar-Gross-Krook (BGK)~\cite{sm-bhatnagar-pr-1954} operator:
\begin{equation}\label{eq:bgk-sm}
  \Omega_i(f) = \frac{\Delta t}{\tau}(f_i^{\rm eq} - f_i) ,
\end{equation}
which relaxes the populations at rate $\tau$ towards a local equilibrium defined by the Maxwell-Boltzmann distribution,
for which we use a second order polynomial expansion:
\begin{equation}\label{eq:feq}
        f^{\rm{eq}}_{i}(\rho, \bm{u})   = \,  w_i \rho 
        \left( 
               1 + \frac{\bm{u} \cdot \bm{c}_{i}}{c_s^2} 
               + \frac{(\bm{u} \cdot \bm{c}_{i})^2 - (c_s |\bm{u}|)^2 }{2c_s^4}
       \right),
  \notag  
\end{equation}
with $c_s = 1 / \sqrt{3}$ the sound speed in the lattice, and $w_i$ a lattice-dependent set of weighting factors
(see Tab.~\ref{tab:stencil-weights}).

Several options are available for implementing the body forces in LBM~\cite{sm-kruger-book-2017}.
In this work we adopt the Exact Difference Method (EDM) scheme~\cite{sm-kupershtokh-cma-2009},
where the external forcing term is computed as 
\begin{equation}\label{eq:fext}
  f_i^{\rm ext} = f_i^{\rm eq, \rm shift} - f_i^{\rm eq} ,
\end{equation} 
where the shifted equilibrium $f_i^{\rm eq, \rm shift}$ is computed from Eq.~\ref{eq:feq} based on the
shifted velocity $\bm{u}^{\text{shift}} = \bm{u} + \frac{F}{\rho}$.

Following an asymptotic analysis, such as the Chapman-Enskog expansion~\cite{sm-chapman-book-1952},
it can be shown that Eq.~\ref{eq:lbe-sm} combined with Eq.~\ref{eq:bgk-sm} delivers in the low-Mach number limit 
a second order approximation of the Navier-Stokes equations, with the following expression putting 
in relationship the relaxation time parameter $\tau$ with the kinematic viscosity $\nu$ of the fluid:
\begin{equation}\label{eq:tau}
  \nu = \left( \tau - \frac{\Delta t}{2} \right) c_s^2 \quad .
\end{equation}

\subsection{Implementing the Smagorinsky Turbulence Model in LBM} 
In the main text we have reported numerical results of LBM simulations equipped with the Smagorinsky SGS model.
The model involves adding an eddy viscosity term to the governing equations to account for 
unresolved turbulent fluctuations at the grid scale. 
This eddy viscosity is computed as:
\begin{equation}\label{eq:smago}
  \nu_{\rm eff} = \nu + \nu_t = \nu + (C_s \Delta)^2 |\overline{\bm{S}}| ,
\end{equation}
where $|\overline{\bm{S}}| = \sqrt{2 \overline{S}_{ij} \overline{S}_{ij}}, \overline{S}_{ij} = 1/2 (\partial_i \overline{u}_{j} + \partial_j \overline{u}_{i} )$ 
is the magnitude of the filtered strain-rate, $\Delta$ is the filter width, and $C_s$ is a 
non-dimensional coefficient called the Smagorinsky constant.

In order to incorporate the Smagorinsky model in a LBM scheme one needs to i) compute the strain rate tensor in each
cell to then ii) compute the eddy viscosity $\nu_{\rm eff}$ to finally iii) locally adjust the relaxation time
to match $\nu_{\rm eff}$.
In LBM the strain rate tensor can be computed locally from the non-equilibrium part of the distribution function:
\begin{equation}
  \overline{S}_{ij} = - \frac{3 \tau}{2} \Pi_{ij}^{\rm neq}, 
  \quad 
  \Pi_{ij}^{\rm neq} = \sum_q c_{q i} c_{q j} \left(f_q - f_q^{\rm eq}  \right) \quad .
\end{equation}
By combining Eq.~\ref{eq:tau} with Eq.~\ref{eq:smago} it is then possible to compute the local relaxation time
due to the Smagorinsky model:
\begin{equation}
  \tau_{\rm eff} = \frac{\nu_{\rm eff}}{c_s^2} + \frac{\Delta t}{2} .
\end{equation}

\section{Artificial Neural Network Architecture and Training Process}

\subsection{Definition of the Training Dataset}
The ground truth data used to train the ANN consists of N pairs of 19-tuples
\begin{equation}\label{eq:training-set}
  \left\{ \left( {f_{i,k}^{\rm pre}}, {f_{i,k}^{\rm post}}  \right), k = 1, 2, \dots , N  \right\} .
\end{equation}
The data is obtained from Direct Numerical Simulation (DNS) of Homogeneous Isotropic Turbulence. 
We conduct LBM simulations in a three-dimensional cubic domain
of side $L = 128$, with periodic boundary conditions and $\Delta t = 1$. 
Simulations are driven by a stationary external force, ensuring a divergence-free flow, 
defined for the $x$, $y$ and $z$ components as:
\begin{equation}
  \begin{cases}
      F_x(y) = F \sin \left(\frac{2 \pi}{L} y \right) \\
      F_y(z) = F \sin \left(\frac{2 \pi}{L} z \right) \\
      F_z(x) = F \sin \left(\frac{2 \pi}{L} x \right)
  \end{cases} .
\end{equation}
We work at a value of the Reynolds number $\rm{Re} \approx 6000$, achieved tuning the relaxation time $\tau$
and the magnitude of the external force as detailed in Tab.~\ref{tab:simulation_parameters}.
We collect data after an initial transient phase of $T_{0}$ timesteps, ensuring that the dynamics reaches a statistically stationary state. %
The full state of the system (in terms of lattice populations) is sampled every $\delta_T$ timesteps, 
for a total of $N_T$ timesteps. 
Next, we define a coarse graining factor $\rm{cg}$ and a filter $\phi$, which allows mapping at a generic time step $t$ 
lattice populations from the $L^3$ DNS grid to the $(L / \rm{cg})^3$ coarse-grained grid:
\begin{equation}
  f^{\rm cg}_i(t) \leftarrow \phi( f^{\rm DNS}_i(t) ) .
\end{equation}
Several possible choices can be made for the filter $\phi$, which may substantially impact the quality of the results.
In this work we consider a sub-sampling strategy, and define the coarse-grained grid by sampling  
one every $\rm{cg}$ lattice points along the different coordinates.
We leave as a future work a careful evaluation and comparison of different choices for $\phi$.

At this stage, we can establish the training set (Eq.~\ref{eq:training-set}) at the coarse grained level, 
following the procedure sketched in Fig.~2 of the main text, which we here summarize:
\begin{enumerate}
  \item We start from the pre-collision populations at the DNS level, $f_{\rm DNS}^{\rm pre}(t)$, and apply a filter 
        in order to obtain $f_{\rm cg}^{\rm pre}(t)$.
  \item We advance the DNS simulation for $\rm{cg} \Delta t$ time-steps obtaining the post-streaming state, i.e.
        $f_{\rm DNS}^{\rm pre}(t + \rm{cg} \Delta t)$.
  \item By filtering the DNS level we obtain $f_{\rm cg}^{\rm pre}(t + \Delta t)$.
  \item In order to obtain the pre-streaming state, i.e. the post-collision values at time $t$, we apply 
        the inverse of the streaming operation at the coarse grained level by reversing the streaming direction, 
        i.e. with respect to $-\bm{c}_i$.
\end{enumerate}

For the results reported in the main text we have used  $\rm{cg} = 2$ and $\rm{cg} = 4$, resulting in 
effective lattice sizes of, respectively, $L^{\prime} = 64$ and $L^{\prime} = 32$. 
Note that to achieve the same $\rm{Re}$ on the coarse grids one could use the same numerical parameters provided in
Tab.~\ref{tab:simulation_parameters}, alongside a rescaling of the relaxation time parameter
\begin{equation}\label{eq:tau-rescaled}
  \tau^{\prime} = \frac{1}{\rm cg} \left( \tau - \frac{1}{2} \right) + \frac{1}{2} \quad .
\end{equation}
However, we shall remark that the choice of the parameters is such that plain LBM simulations at the coarse grid level
would lead to numerical instabilities.

\begin{table}[ht]
  \centering
  \begin{tabular}{c | p{4cm} | p{2cm}}
    \hline
    \textbf{Parameter} & \textbf{Description}                & \textbf{Value} \\
    \hline
    \( \tau \)         & Relaxation time Parameter           & $0.5032$ \\
    \( \nu \)          & Viscosity                           & $1.07 \times 10^{-3}$ \\
    \( F \)            & Force term Amplitude                & $5 \times 10^{-6}$ \\
    \( \Delta_x \)     & Grid size                           & $1$ \\
    \( \Delta_t \)     & Timestep                            & $1$ \\
    \( L \)            & Lattice Size for DNS                & $128$ \\
    \( \mathbf{u}_{RMS} \) & RMS of velocity       & $5 \times 10^{-2}$ \\
    \( \rm{Re} \)      & Reynolds Number                     & $6 \times 10^3$ \\
    \( \lambda \)      & Taylor microscale                   & $5.256$ \\
    \( (\mathbf{v}')_{RMS} \)   & RMS of velocity fluctuations   & $1.567e-2$ \\
    \( Re_{\lambda} \) & Taylor m. Reynolds Number            & $77.45$ \\
    \( T_L \)          & Large scale char. time     & $2.560 \times 10^3$ \\
    \( T_{\lambda} \)  & Taylor scale char. time     & $3.354 \times 10^2$ \\
    \hline
    \( T_0 \)          & Transient Phase Duration            & $5 \times 10^{4}$ \\
    \( \delta_T \)     & Sampling Interval                   & $10^{3}$ \\
    \( N_T \)          & Total Number of Timesteps           & $10^{3}$ \\
    \( M \)            & Points Sampled per Tstep & $512$ \\
    \hline
    \( \lambda \)      & Learning rate                       & $1 \times 10^{-4}$ \\
    \( \omega \)       & Weight decay rate                   & $1 \times 10^{-6}$ \\
    \hline
  \end{tabular}
  \caption{ The table reports the numerical parameters used in DNS numerical simulations (top),
            for the definition of the training dataset (middle) and for the hyper-parameters
            used by the ADAM optimizer to train the ANN.
          }
  \label{tab:simulation_parameters}
\end{table}

\subsection{Neural Network Architecture and Optimization}
In this section we detail our ANN model, which consists in learning a correction term adding to the BGK collision operator.
We define the post-collision state as
\begin{equation}\label{eq:nn-model-full}
  \tilde{f}_i^{\rm post} = f_i^{\rm pre} + \Omega(f_i^{\rm pre}) + \frac{1}{\tau} \Omega^{\rm NN}(f_i^{\rm pre})
\end{equation}
where $\Omega$ is the BGK collision operator in Eq.~\ref{eq:bgk-sm}, while $\Omega^{\rm NN}$ is the
correction term due to the ANN.

We consider an ANN consisting of a Multi-layer Perceptron (MLP) with input and output layers of size $19$
(equal to the number of lattice populations sitting on one single grid point), and three fully connected hidden layers 
of size $[300, 300, 100]$, in combination with LeakyReLU activation function.
Following Ref.~\cite{sm-corbetta-epje-2023}, before the output layer we also include a non-trainable layer 
posing hard constraints on the conservation of mass and momentum. In our case we require that the correction 
term given by the ANN does not introduce additional mass and momentum in the output:
\begin{align}\label{eq:nn-conservation}
  \sum_i \Omega^{\rm NN}(f^{\rm pre}_i)            &= 0       , \\
  \sum_i \Omega^{\rm NN}(f^{\rm pre}_i) \bm{c}_{i} &= \bm{0}  .
\end{align}
The above equations are satisfied using
\begin{align}
  \hat{f}_i &= \Omega^{\rm NN}(f^{\rm pre}_i) + \kappa_1 + \kappa_2 c_{i,x} + \kappa_3 c_{i,y} + \kappa_4 c_{i,z} ,
\end{align}
where $\hat{f}_i$ represents the adjusted output ensuring mass and momentum conservation,
and with the following coefficients specific to the D3Q19 model:
\begin{align*}
  \kappa_1  &= - \frac{1}{19} \sum_i \Omega^{\rm NN}(f^{\rm pre}_i)         , \\
  \kappa_2  &= - \frac{1}{10} \sum_i \Omega^{\rm NN}(f^{\rm pre}_i) c_{i,x} , \\
  \kappa_3  &= - \frac{1}{10} \sum_i \Omega^{\rm NN}(f^{\rm pre}_i) c_{i,y} , \\
  \kappa_4  &= - \frac{1}{10} \sum_i \Omega^{\rm NN}(f^{\rm pre}_i) c_{i,z} .
\end{align*}

We have observed that the model defined in Eq.~\ref{eq:nn-model-full}, albeit performing well in one-step predictions,
once plugged in a simulation typically leads to numerical instabilities after just few iterations.
This behavior is commonly observed in autoregressive models, i.e. models that feed in the output as an input 
for the next step~\cite{sm-brandstetter-arxiv-2022}. This is due to accumulations of errors 
that eventually deviate the dynamics of the system towards configurations not covered in the training dataset, 
a phenomenon known in literature as \textit{distribution shift} \cite{sm-goodfellow-book-2016}.
In the next section we discuss a procedure which allows to mitigate this problem.

\subsection{Unrolled training}
In this section we discuss the \textit{Unrolled training} technique, which we employ to enhance the stability of the model. 
In an auto-regressive model, i.e. a model that recursively uses its output as an input for the successive timesteps, 
one must ensure the stability of the model with respect to the accumulation of errors. 
This is a requirement, for example, for any time integration scheme, where the stability hinges on both the system's 
dynamics and the discretization scheme employed. 
Moreover, in the domain of Machine Learning, another phenomenon may arise, since the accumulation of errors 
may lead the dynamics of the system into unexplored regions of the optimization space, deviating from the training conditions.
This problem is typically framed in the context of \textit{distribution shift}, and deals with scenarios 
in which  the data distribution during training differs from that encountered during testing. 
Fields such as language modeling and robotics commonly encounter such challenges. 
For instance, maintaining coherence in generated text over multiple iterations is a complex task
for language models~\cite{sm-holtzman-arxiv-2019}.

Several solutions have been proposed to deal with this problem. One possibility is to modify the architecture of 
the Neural Network, for example forcing the spectra of the linear matrices, stabilizing the dynamics. 
Examples of this approach are Antisymmetric Neural Networks \cite{sm-goles-dam-1986}. 
Another possibility is to use regularization strategies such as noise injection in the input and/or 
the output of the Neural Network \cite{sm-labach-arxiv-2019}. 

In this work, we employ a technique known in literature as \textit{unrolled training} 
or \textit{trajectory learning} \cite{sm-brandstetter-arxiv-2022}. 
In short, it consists in applying the model to make predictions over $L_T$ consecutive timesteps at training time,
backpropagating the gradients through some or all such timesteps. 
This is similar to the \textit{Backpropagation Through Time} method, used for training Recurrent Neural Networks, 
with the main difference being that in \textit{unrolled training} no memory term is present in the Neural Network. 

We train the model using the following error metric:
\begin{equation}\label{eq:nn-loss}
  \mathcal{L} = \sum_{x_i \in S} \left( \tilde{f}^{\rm post}(x_i, t_0 + L_t) - f^{\rm post}(x_i, t_0 + L_t) \right)^2 ,
\end{equation}
where for each training epoch $S$ represents a randomly selected subset of the lattice points.
One single training epoch consists of the following steps:
\begin{enumerate}
    \item We select a snapshot from the ground truth dataset, i.e. the pre-collision populations on the entire filtered
          grid of size $(L/\rm{cg})^3$ at a fixed time step. This serves as the initial condition.
    \item We randomly select a subset of lattice points, denoted as $S=\{ x_i \}_{i=0}^{batch\_size}$, which 
          will be used compute the loss.
    \item We evolve the system for $L_T$ timesteps, alternating the collision step via Eq.~\ref{eq:nn-model-full}, 
          and the streaming step.
    \item We compute the loss via Eq.~\ref{eq:nn-loss} comparing against ground-truth values at time $t + L_T$.
    \item Finally, we backpropagate the loss through all the timesteps to compute the gradients 
          and update the weights of the Neural Network.
\end{enumerate}
We remark that, while the full domain is evolved using the ANN, the loss term is computed
only on randomly selected subset of points. The reason for this is twofold. First, it allows reducing 
computational costs, both in terms of memory requirements as well as overall training time. Second, it allows to increase
stochasticity in the computation of gradients for updating the weights of the network.

This strategy introduces a few extra hyperparameters in the training of the network.
The results reported in the main text have been obtained using a batch size $|S|$ of $512$ elements, with a trajectory 
lenght of $L_T = 24$ for $cg=2$ and $L_T = 12$ for $cg=4$.
The choice of these parameters impacts the time and memory requirements of the training process
(longer trajectories and bigger batch sizes require more time and memory), and on the resulting stability 
of the model (too short trajectories give less stable models).
The optimization is done using Adam Optimizer, training for a total of $2 \times 10^4$ epochs with early 
stopping based on test loss. The optimization parameters are reported in Table \ref{tab:simulation_parameters}.

\section{Structure Functions and Extended Self Similarity}

In Fig.~4 in the main text we have reported results with of high order statistical properties of turbulence, analyzing
data from simulations using our data-driven SGS model.
In this section we provide further details; for the convenience of the reader we state once again the definition of
the Eulerian Structure function of order $p$:
\begin{equation}\label{eq:structure-fun-sm}
  S^p(l) = \left\langle \left[\left( \bm{u}(\bm{x} + \bm{l}) - \bm{u}(\bm{x}) \right)\cdot {\hat{\bm{l}}}\right]^p \right\rangle .
\end{equation}
\begin{figure}[htb]
  \centering
  \begin{overpic}[width=.99\textwidth]{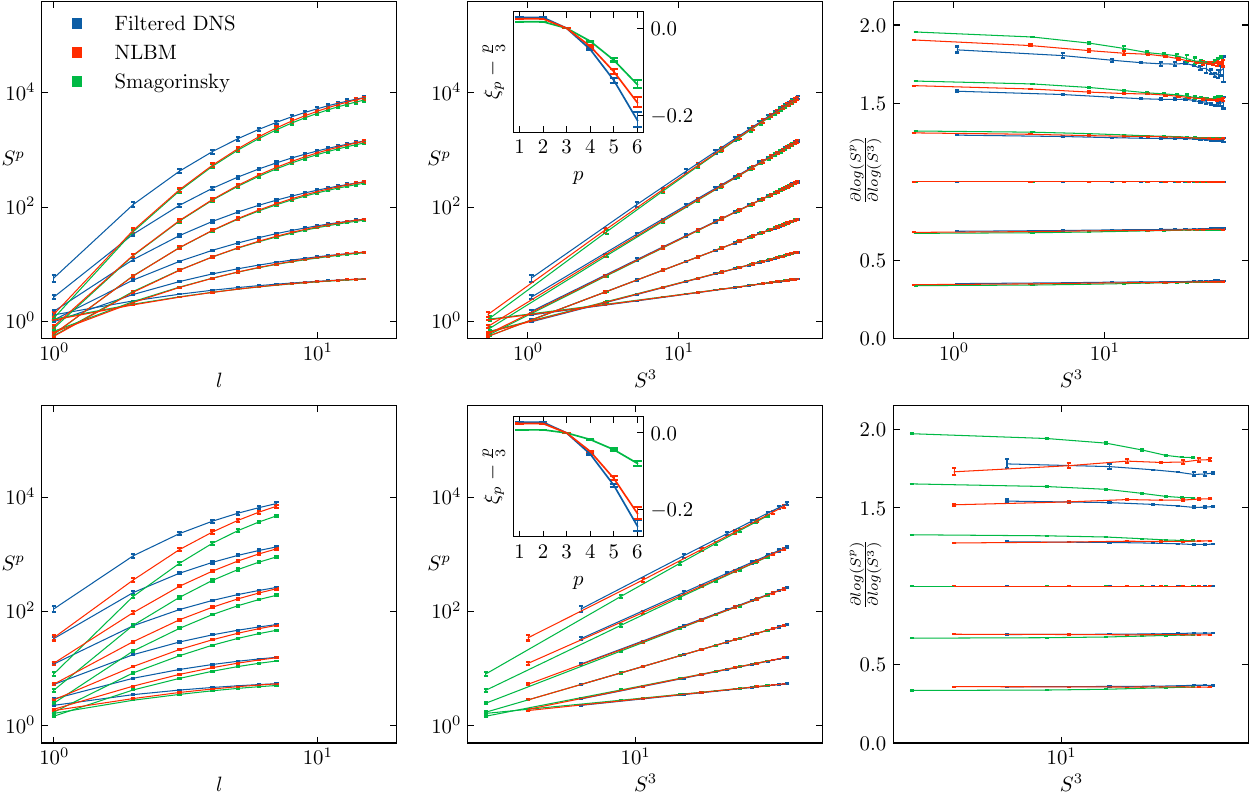}
    \put( 0 , 63.0){(a)}
    \put(34 , 63.0){(b)}
    \put(68 , 63.0){(c)}
    \put( 0 , 30.5){(d)}
    \put(34 , 30.5){(e)}
    \put(68 , 30.5){(f)}  
    \put(25 , 61.0){$\rm{cg} = 2$}
    \put(59 , 61.0){$\rm{cg} = 2$}
    \put(93 , 61.0){$\rm{cg} = 2$}
    \put(25 , 28.5){$\rm{cg} = 4$}
    \put(59 , 28.5){$\rm{cg} = 4$}
    \put(93 , 28.5){$\rm{cg} = 4$}
  \end{overpic}  
  \caption{Results for structure functions (cf. Eq.~\ref{eq:structure-fun-sm}) of order $p$, ranging between $p = 1$ to $p = 6$, 
           with in blue data from filtered DNS, in red data from simulations using NLBM, 
           and in green data from simulations using the Smagorinsky model.
           On the top row (panels (a)-(c)) results for $cg=2$, on the bottom row (panels (d)-(f)) results for $cg=4$. 
           In the first column we report the Structure function $S^p(l)$ versus $l$, in the second column 
           $S^p$ versus $S^3$, and in the third column the logarithmic derivative $\frac{\partial log(S^p)}{\partial log(S^3)}$. 
          }\label{fig:ess-extended}
\end{figure}
In Fig.~\ref{fig:ess-extended}, the first row refers to results for a grid with coarse graining factor $\rm{cg} = 2$,
whereas the second row shows results for $\rm{cg} = 4$.
For all cases we show in blue color filtered DNS data, in red our NLBM model, and for comparison in green results
using Smagorinsky.
In the first column we plot the structure functions $S^p$ vs $l$, with $p$ ranging between $1$ to $6$.
The central column shows the results of the extended self similarity analysis, where this time we plot the structure
functions $S^p$ versus versus $S^3$. The insets shows the deviation of the scaling exponents $\xi_p$ from K41
scaling $p/3$. The scaling exponents have been computed fitting the slopes, which are better highlighted in the third 
column, where we show the logarithmic derivative $\frac{\partial log(S^p)}{\partial log(S^3)}$. 
The fitting ranges are defined considering intervals where the logarithmic derivative stays approximately constant, 
avoiding the dissipative interval by excluding the first point, resulting in $S^3$ values of $[4,20]$ for $cg=2$ and $[8,40]$ for $cg=4$.

The results here presented strengthen the message and the considerations made in the main text, 
showcasing that NLBM better captures the kinetic energy at the large scales with respect than Smagorinsky,
which exhibits instead larger dissipation.

\section{Generalization}

A primary concern in data-driven modelling is over-fitting and poor generalization \cite{sm-goodfellow-book-2016}, 
connected to the challenge of generating a comprehensive, high-quality training dataset that can cover all potential uses. 
Additionally, ANN models often lack interpretability and physical consistency, rendering them unreliable in application contexts.
In this section, we discuss the generalization capabilities of our model. 

\begin{figure}[htb]
  \centering
  \begin{overpic}[width=.99\textwidth]{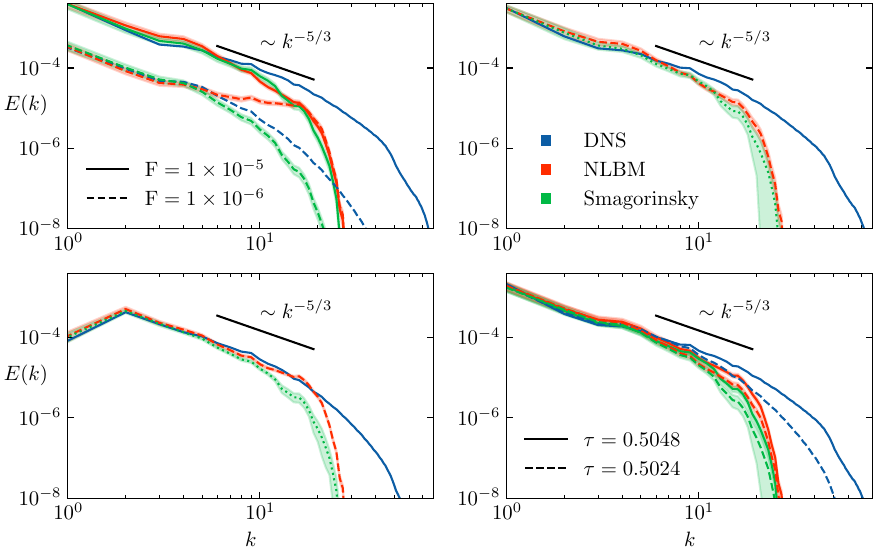}
    \put(0 ,61){(a)}
    \put(0 ,30){(c)}
    \put(52,30){(d)}
    \put(52,61){(b)}
  \end{overpic}  
  \caption{ Energy spectrum for simulations of HIT with different parameters to evaluate the capability of NLBM 
            to generalize outside of the training dataset.
            Blue curves represent DNS results, red for NLBM and green for the Smagorinsky SGS.
            In panel a) we present results increasing or decreasing the magnitude of the forcing term,
            giving respectively $Re \approx 1.2 \times 10^4$ and $Re \approx 1.2 \times 10^3$ (vs $Re \approx 6 \times 10^3$ used at training time).
            In panel b) we show results using a non-homogeneous forcing term.
            In panel c) we show the effect of forcing with a higher wave number.
            In panel d) we show the effect of increasing and decreasing the relaxation time parameter (hence the kinematic viscosity).
          }\label{fig:spectra-generalized}
\end{figure}

We consider the ANN trained with the parameters from Tab.~\ref{tab:simulation_parameters}, with $\rm{cg} = 4$, 
and consider a few scenarios slightly departing from the training conditions.
In Fig.~\ref{fig:spectra-generalized} we report the energy spectrum for four different cases,
comparing the results of DNS simulations, NLBM and Smagorinsky, changing the following parameters.
Panels from (a) to (d) cover the following set of parameters:
\begin{enumerate}[(a)]
  \item Different magnitude of the forcing term (for reference, value used during training: $F=5 \times 10^{-6}$, $Re \approx 6 \times 10^3$) %
  \begin{itemize}
      \item  $F = 1 \times 10^{-5}$ ( $Re \approx 1.2 \times 10^4$) ,
      \item  $F = 1 \times 10^{-6}$ ( $Re \approx 1.2 \times 10^3$) .
  \end{itemize}
  \item A non-homogeneous forcing term:
  \begin{equation*}
    \begin{cases}
        F_x(y) = 2 F \sin \left(\frac{2 \pi}{L} y \right) \\
        F_y(z) = 0 \\
        F_z(x) = 0
    \end{cases} .
  \end{equation*}
  \item A higher wave number in the \textit{sine} function:
  \begin{equation*}
    \begin{cases}
        F_x(y) = F \sin \left(2 \frac{2 \pi}{L} y \right) \\
        F_y(z) = F \sin \left(2 \frac{2 \pi}{L} z \right) \\
        F_z(x) = F \sin \left(2 \frac{2 \pi}{L} x \right) 
    \end{cases} .
  \end{equation*}
  \item Higher and lower values of the kinematic viscosity, changing the relaxation time (for reference, value used in training $\tau=0.5032$, $Re \approx 6 \times 10^3$)
\end{enumerate}
  \begin{itemize}
      \item  $\tau=0.5048$ ( $Re \approx 4 \times 10^3$) ,
      \item  $\tau=0.5024$ ( $Re \approx 8 \times 10^3$) .
  \end{itemize}

The results show that the NLBM yields stable simulations for all the cases covered, considering both higher 
and lower value for $\rm{Re}$ with respect to the one considered for training the ANN. 
NLBM provides results systematically in better agreement with DNS with respect to Smagorinsky,
with the exception of one case in Fig.~\ref{fig:spectra-generalized}(a) where the absolute value of the external force
was taken to be 5 times smaller than the one used during training.
We attribute this to a possible overfitting of the small scales, which, as can be seen in the plot, overlap for NLBM
regardless of the magnitude of the external force.

We should also remark that the generalization capabilities of the model do not extend to different grid sizes 
and different coarse graining factor; such cases currently require the training of a new model from scratch.

\section{Physical Interpretation of model action}
In this section we provide a physical interpretation of the action of the ANN.
While our model works at the kinetic level, it is simple to map the lattice populations $f$ to the moments space,
and observe the action of the model in terms of physical quantities.
In order to outline the procedure, we start from Eq.~\ref{eq:lbe-sm} with the single-relaxation time BGK operator 
in Eq.~\ref{eq:bgk-sm}, introduce an invertible matrix $\bm{M}$, and recast the equation in the following form: 
\begin{equation}\label{eq:mrt}
  \begin{split}
      f^{\rm post} - f^{\rm pre} = \bm{M}^{-1} \frac{1}{\tau} \bm{M} \left(f^{\rm eq} - f^{\rm pre} \right) + f^{\rm ext}\\
                                 = \bm{M}^{-1}         \bm{S} \bm{M} \left(f^{\rm eq} - f^{\rm pre} \right) + f^{\rm ext}
  \end{split} ,
\end{equation}
where $\bm{M}$ defines a transformation from lattice populations $f(\bm{x}, t)$ to macroscopic moments 
$\bm{m} = \bm{M} f(\bm{x}, t)$, and the relaxation matrix 
$\bm{S} = \text{diag}\left( \frac{1}{\tau}, \dots,  \frac{1}{\tau} \right)$ acts on macroscopic moments space. 

A generalization of the BGK collision operator in given by the \textit{Multi-Relaxation Time} (MRT) 
collision operator~\cite{sm-dhumieres-ptsa-2002}, which allows for individual relaxation rates 
for the different macroscopic moments.
\begin{figure}[tb]
  \centering
  \begin{overpic}[width=.50\textwidth]{{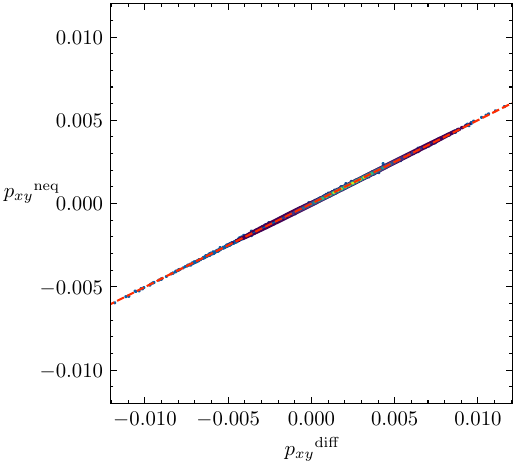}}
    \put(2 ,87){(a)}
  \end{overpic}
  \begin{overpic}[width=.49\textwidth]{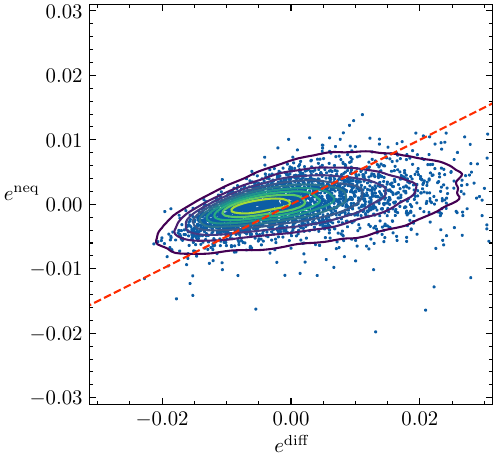}
    \put(2, 90){(b)}  
  \end{overpic}      
  \caption{
            Scatterplots of $\bm{m^{\rm neq}}$ versus $\bm {m^{\rm diff}}$ (see text for their definition) from NLBM data 
            for the components corresponding to the component $p_{xy}$ of the shear stress tensor in (a) and to
            the kinetic energy $e$ in (b).
            The data shown here comes from a model trained using a coarse graining factor $cg=4$. 
            In both panels, the red dotted line correspond to the value used for the BGK model at training time 
            $\tau^{\prime} = 0.5008$ (rescaled with respect to $\rm{cg}$ as per Eq.~\ref{eq:tau-rescaled}).
            A clear linear dependence can be observed for case $p_{xy}$, which allows to define an effective 
            relaxation time $\tau_{\rm eff} \approx 0.5023 $, slightly larger than $\tau^{\prime}$, hence leading
            to a larger effective value of the kinetic viscosity.
            In panel b) we see instead strong non-linear effects which does not allow for fitting a value of the 
            bulk viscosity.
          }\label{fig:moments-dist-mrt}
\end{figure}
In this framework, several possible choices can be operated for the matrix $\bm{M}$, corresponding to a 
map of the lattice populations into different sets of macroscopic moments $\bm{m}$, in which the individual relaxation times 
$\tau_1$, ..., $\tau_{19}$ can have different physical interpretations. 
In what follows we consider
\setcounter{MaxMatrixCols}{20}
\begin{equation}
  \bm{M} =     
  \begin{pmatrix}
      1 & 1 & 1 & 1 & 1 & 1 & 1 & 1 & 1 & 1 & 1 & 1 & 1 & 1 & 1 & 1 & 1 & 1 & 1 \\
    -30 & -11 & -11 & -11 & -11 & -11 & -11 & 8 & 8 & 8 & 8 & 8 & 8 & 8 & 8 & 8 & 8 & 8 & 8 \\
     12 & -4 & -4 & -4 & -4 & -4 & -4 & 1 & 1 & 1 & 1 & 1 & 1 & 1 & 1 & 1 & 1 & 1 & 1 \\
      0 & -1 & 0 & 0 & 0 & 0 & 1 & -1 & -1 & -1 & -1 & 0 & 0 & 0 & 0 & 1 & 1 & 1 & 1 \\
      0 & 4 & 0 & 0 & 0 & 0 & -4 & -1 & -1 & -1 & -1 & 0 & 0 & 0 & 0 & 1 & 1 & 1 & 1 \\
      0 & 0 & -1 & 0 & 0 & 1 & 0 & -1 & 0 & 0 & 1 & -1 & -1 & 1 & 1 & -1 & 0 & 0 & 1 \\
      0 & 0 & 4 & 0 & 0 & -4 & 0 & -1 & 0 & 0 & 1 & -1 & -1 & 1 & 1 & -1 & 0 & 0 & 1 \\
      0 & 0 & 0 & -1 & 1 & 0 & 0 & 0 & -1 & 1 & 0 & -1 & 1 & -1 & 1 & 0 & -1 & 1 & 0 \\
      0 & 0 & 0 & 4 & -4 & 0 & 0 & 0 & -1 & 1 & 0 & -1 & 1 & -1 & 1 & 0 & -1 & 1 & 0 \\
      0 & 2 & -1 & -1 & -1 & -1 & 2 & 1 & 1 & 1 & 1 & -2 & -2 & -2 & -2 & 1 & 1 & 1 & 1 \\
      0 & -4 & 2 & 2 & 2 & 2 & -4 & 1 & 1 & 1 & 1 & -2 & -2 & -2 & -2 & 1 & 1 & 1 & 1 \\
      0 & 0 & 1 & -1 & -1 & 1 & 0 & 1 & -1 & -1 & 1 & 0 & 0 & 0 & 0 & 1 & -1 & -1 & 1 \\
      0 & 0 & -2 & 2 & 2 & -2 & 0 & 1 & -1 & -1 & 1 & 0 & 0 & 0 & 0 & 1 & -1 & -1 & 1 \\
      0 & 0 & 0 & 0 & 0 & 0 & 0 & 1 & 0 & 0 & -1 & 0 & 0 & 0 & 0 & -1 & 0 & 0 & 1 \\
      0 & 0 & 0 & 0 & 0 & 0 & 0 & 0 & 0 & 0 & 0 & 1 & -1 & -1 & 1 & 0 & 0 & 0 & 0 \\
      0 & 0 & 0 & 0 & 0 & 0 & 0 & 0 & 1 & -1 & 0 & 0 & 0 & 0 & 0 & 0 & -1 & 1 & 0 \\
      0 & 0 & 0 & 0 & 0 & 0 & 0 & -1 & 1 & 1 & -1 & 0 & 0 & 0 & 0 & 1 & -1 & -1 & 1 \\
      0 & 0 & 0 & 0 & 0 & 0 & 0 & 1 & 0 & 0 & -1 & -1 & -1 & 1 & 1 & 1 & 0 & 0 & -1 \\
      0 & 0 & 0 & 0 & 0 & 0 & 0 & 0 & -1 & 1 & 0 & 1 & -1 & 1 & -1 & 0 & -1 & 1 & 0 \\
  \end{pmatrix} ,
\end{equation}
which maps to the following macroscopic moments:
\begin{equation}
  \bm{m} = 
  \begin{pmatrix}
    \rho, & e, & \epsilon, & j_x, & q_x, & j_y, & q_y, & j_z, & q_z, & p_{xx}, & \pi_{xx}, & p_{ww}, & \pi_{ww}, & p_{xy}, & p_{yz}, & p_{xz}, & m_{x}, & m_{y}, & m_{z}
  \end{pmatrix} ,
\end{equation}
where:
\begin{itemize}
  \item $\rho$, $j_x$, $j_y$ and $j_z$ are the mass and the components of momentum along $x,y,z$.
  \item $e$ is the kinetic energy;
  \item $\epsilon$ is the kinetic energy squared;
  \item  $q_x$, $q_y$ and $q_z$ are the heat fluxes along $x,y,z$.
  \item $p_{xx}$, $p_{ww}$, $p_{xy}$, $p_{yz}$ and $p_{xz}$ are the components of the symmetric traceless viscous stress tensor;
  \item $\pi_{xx}$, $\pi_{ww}$, $m_x$, $m_y$ and $m_z$ correspond to higher order moments, 
        with no obvious physical interpretation (in the LBM jargon they go under the name of ghost modes).
\end{itemize}
In turn, the macroscopic moments are associated to the following relaxation parameters:
\begin{equation}\label{eq:relax-mrt}
  \bm{S} = \rm{diag}
  \begin{pmatrix}
    0, & \tau_e, & \tau_{\epsilon}, & 0, & \tau_q, & 0, & \tau_q, & 0, & \tau_q, & \tau_{\nu}, & \tau_{\pi}, & \tau_{\nu}, & \tau_{\pi}, & \tau_{\nu}, & \tau_{\nu}, & \tau_{\nu} , & \tau_m, & \tau_m, & \tau_m
  \end{pmatrix} ,
\end{equation}
where the relaxation times associated to conserved quantities have been set to zero.
\begin{figure}[tb]
  \centering
  \begin{overpic}[width=.99\textwidth]{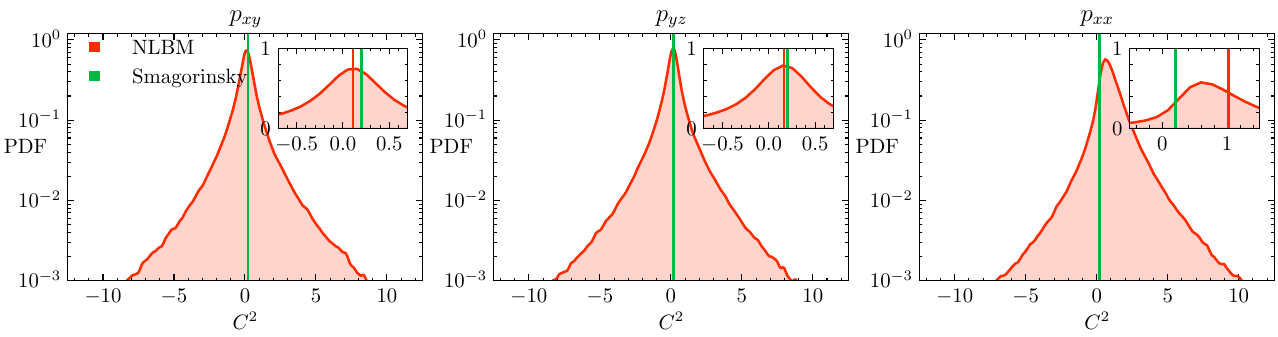}
    \put(1 , 25){(a)}
    \put(34, 25){(b)}
    \put(68, 25){(c)}
  \end{overpic}
  \caption{
            Probability distribution function (PDF) of the fitted
            value of the Smagorinsky constant $C^2$ from NLBM data, for coarse graining factor $cg=4$.
            In panel a) data fitted considering the $p_{xy}$ component of the shear stress tensor, 
            in b) using $p_{yz}$ and in c) using $p_{xx}$.
            The inset highlights that the average value, marked with vertical bars, comparing 
            the results of NLBM (red) against the value used in the Smagorinsky SGS model in green ($C^2 = 0.2$).
          }\label{fig:nu-eff-extended}
\end{figure}

Likewise for the BGK, also in MRT~\cite{sm-dhumieres-ptsa-2002} it is possible to establish a connection between
the relaxation time parameters and the macroscopic tranport coefficients. For example, the kinematic 
kinematic viscosity $\nu$ and bulk viscosity $\zeta$ of the model are given by:
\begin{equation}
  \nu = c_s^2 \left( \tau_{\nu} - \frac{1}{2}\right); \:\:\: \zeta = \frac{5 - 9 c_s^2}{9}\left( \tau_e - \frac{1}{2} \right) , %
\end{equation}

While for NLBM an asymptotic analysis for defining the transport coefficients is not viable from an analytic point 
of view, it is still possible to numerically evaluate how the ANN relaxes the different moments $\bm{m}$. 
To this aim we multiply the LHS and the RHS of Eq.~\ref{eq:mrt} by $\bm{M}$:
\begin{equation}
  \begin{split}
             \underbrace{ \bm{M} \left( f^{\rm post} - f^{\rm pre} - f^{\rm ext} \right) }_{\bm{m^{\rm diff}}} 
      = 
      \bm{S} \underbrace{ \bm{M} \left( f^{\rm eq  } - f^{\rm pre} \right) }_{\bm{m^{\rm neq  }}}
  \end{split} ,
\end{equation}
and define $\bm{m^{\rm diff}}$ and $\bm{m^{\rm neq}}$, two column vectors defined in the moment 
space similarly to $\bm{m}$.

From numerical data, we can plot the different components of $\bm{m^{\rm diff}}$ vs $\bm{m^{\rm neq  }}$.
In Fig.~\ref{fig:moments-dist-mrt} we report and example of this analysis, for the case $\rm{cg} = 4$, 
where the individual components are centered with respect to their average value.
We remark that in a MRT collision operator we would observe a linear dependence, %
with a slope of the regression line $1 / \tau_i$ depending on the choice of the relaxation parameters in $\bm{S}$.
For NLBM we observe, to good approximation, a linear relationship for the moments related to the shear stress tensor
$p_{ij}$. In Fig.~\ref{fig:moments-dist-mrt}(a) we provide an example for the component $p_{xy}$.
From a linear fit we can obtain an effective value for the relaxation time $\tau_{\rm eff}$.
At variance with Eq.~\ref{eq:relax-mrt}, we observe two different relaxation times, one for the components
$p_{xx}$ and $p_{ww}$ ($\tau_{\rm eff} \approx 0.5125$), and a second one for $p_{xy}$, $p_{yz}$ and $p_{xz}$,
($\tau_{\rm eff} \approx 0.5023$). Both values are larger than the corresponding value of the relaxation time 
rescaled as for Eq.~\ref{eq:tau-rescaled}, $\tau = 0.5008$, thus implying a larger effective shear viscosity.

For all the other moments we generally observe a non linear dependence. 
In Fig.~\ref{fig:moments-dist-mrt}(b) we show an example for the kinetic energy coefficient which relates to the 
bulk viscosity.
It is common practice in literature to adjust the bulk viscosity to enhance the stability of numerical methods,
and recently an ANN-based approach has been reported also for LBM~\cite{sm-horstmann-cf-2024}.

To further test the importance of non-linearities introduced by the ANN we have tested a MRT collision operator
in which the matrix $\bm{S}$ has been obtained performing a linear fit of $\tau_{\rm eff}$ for all the different moments
independently (hence discarding the non-linear contributions): %
\begin{equation}
  \bm{S} = \rm{diag} 
  \begin{pmatrix}
      0,  & \tau_1   , & \tau_2   , & 0,         & \tau_3   , & 0,         & \tau_4   , & 0,  & \tau_5   , & \tau_6   , 
          & \tau_7   , & \tau_8   , & \tau_9   , & \tau_{10}, & \tau_{11}, & \tau_{12}, 
          & \tau_{13}, & \tau_{14}, & \tau_{15},
  \end{pmatrix}
\end{equation}
with 
\begin{equation*}
  \begin{split}
  \tau_1   &= 0.1554, \quad 
  \tau_2    = 0.0947, \quad 
  \tau_3    = 0.5220, \quad 
  \tau_4    = 0.6921, \quad 
  \tau_5    = 0.5614, \quad \\
  \tau_6   &= 0.5122, \quad 
  \tau_7    = 0.5564, \quad 
  \tau_8    = 0.5121, \quad 
  \tau_9    = 0.5483, \quad 
  \tau_{10} = 0.5020, \quad \\
  \tau_{11}&= 0.5028, \quad 
  \tau_{12} = 0.5023, \quad 
  \tau_{13} = 0.4597, \quad 
  \tau_{14} = 0.4910, 
  \tau_{15} = 0.4872
  \end{split}
\end{equation*}

With this choice of $\bm{S}$ numerical simulations became unstable after just few iterations, similarly to the BGK case.
We leave as a future work further investigation on the role of the heat flux and of the ghost modes.
We focus here instead on the effective viscosity, which can be computed from the shear stress tensor
akin to the Smagorinsky turbulence model.
From the linear fit of the components of the shear stress tensor it is possible to compute $\nu_{\rm eff}$,
which in combination with Eq.~\ref{eq:smago} allows to define what would be the equivalent of the
Smagorinsky constant for NLBM.
In Fig.~\ref{fig:nu-eff-extended} we report the PDF of the fitted value of the Smagorinsky constant from
NLBM data, with in panel (a) and (b) the results for $p_{xy}$ and $p_{yz}$ respectively, and in $(c)$ for $p_{xx}$.
The average values are comparable with Smagorinsky for cases (a) and (b), about a factor two smaller than 
the one used in simulations with the Smagorinsky SGS model ($C^2 = 0.2$), with the largest discrepancies 
observed for case (c).
For all cases, the presence of a tail with negative values highlights the fact that in NLBM it is 
possible to capture the inverse transfer of energy from small to large scales.

\begin{figure}[tb]
  \centering
  \includegraphics[width=.4\textwidth]{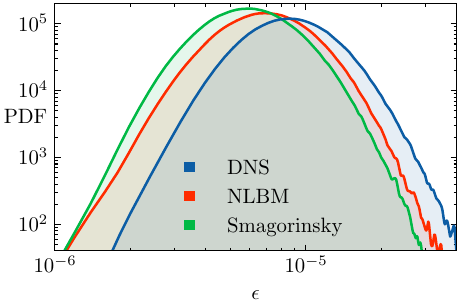}
  \caption{
          Probability distribution function (PDF) of the dissipation field $\epsilon$, computed according to Eq.~\ref{eq:diss}, 
          comparing data from DNS (blue), against NLBM (red) and the Smagorinsky model (green).
          Results for NLBM and the Smagorinsky model are based on simulations using a coarse-graining factor $\rm{cg} = 4$.
          }\label{fig:diss}
\end{figure}

Finally, Fig.~\ref{fig:diss} shows the probability distribution function of the dissipation field 
for the Smagorinsky model, DNS, and NLBM. The dissipation field is computed using:
\begin{equation}\label{eq:diss}
  \epsilon = 2\nu S_{ij}S_{ij} 
\end{equation}
where \( \nu \) is the kinematic viscosity and \( S_{ij} \) is the strain rate tensor. 

The figure presents results for a coarse-graining factor $\rm{cg} = 4$, and shows once more that our model 
can correctly reproduce the small-scale dynamics of HIT, 
replicating the characteristics of the dissipation field as observed in the DNS data.

\end{document}